\documentstyle[11pt,paspconf,epsf]{article}
\begin{document}

\title{Anomalous Microwave Emission}
\author{A. Kogut}
\affil{Laboratory for Astronomy and Solar Physics,
Code 685,
Goddard Space Flight Center,
Greenbelt, MD 20771}

\begin{abstract}
Improved knowledge of diffuse Galactic emission
is important to maximize the scientific return
from scheduled CMB anisotropy missions.
Cross-correlation of microwave maps
with maps of the far-IR dust continuum
show a ubiquitous microwave emission component
whose spatial distribution is traced by far-IR dust emission.
The spectral index of this emission,
$\beta_{\rm radio} = -2.2^{+0.5}_{-0.7}$,
is suggestive of free-free emission
but does not preclude other candidates.
Comparison of H$\alpha$ and microwave results
show that both data sets have positive correlations
with the far-IR dust emission.
Microwave data, however, are consistently brighter
than can be explained solely from free-free emission
traced by H$\alpha$.
This ``anomalous'' microwave emission can be explained as
electric dipole radiation from small spinning dust grains.
The anomalous component at 53 GHz
is 2.5 times as bright
as the free-free emission traced by H$\alpha$,
providing an approximate normalization
for models with significant spinning dust emission.

\end{abstract}

% Keywords should be included, but they are not printed in the hardcopy.
\keywords{ISM, dust, free-free}

\section{Introduction}

Observations of anisotropy in the cosmic microwave background
are complicated by the presence of foreground Galactic emission
along all lines of sight.
At high latitudes ($|b| > 20\deg$),
diffuse Galactic emission is dominated by
optically thin synchrotron, dust, and free-free emission.
In principle, these components may be distinguished by their different
spatial morphology and frequency dependence.
In practice, there is no emission component for which
both the frequency dependence and spatial distribution are well determined.
Synchrotron radiation dominates radio-frequency surveys,
but the spectral index
steepens with frequency and
has poorly-determined spatial variation
(Banday \& Wolfendale 1991,
Bennett et al.\ 1992).
Dust emission dominates far-infrared surveys,
but its spectral behavior at longer wavelengths
depends on the shape, composition, and size distribution
of the dust grains, which are poorly known
(D\'{e}sert, Boulanger, \& Puget 1990).
Free-free emission from electron-ion interactions
has well-determined spectral behavior
but lacks an obvious template map:
free-free emission never dominates the high-latitude radio sky,
while other tracers of the warm ionized interstellar medium (WIM)
such as H$\alpha$ emission, pulsar dispersion measure,
or N II emission
are either incomplete, undersampled, or noise-dominated
(Reynolds 1992;
Fixsen, Bennett, \& Mather 1999).

% --------------- Figure 1: Spectrum of correlated emission --------------- 
\begin{figure}[t]
\plotone{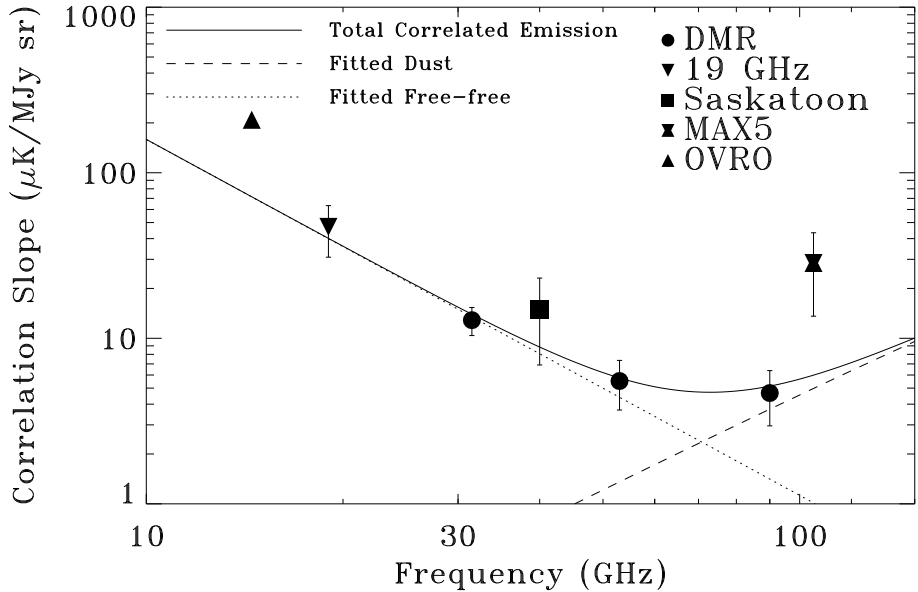}
\caption{Spectrum of correlated intensity fluctuations
between microwave and far-infrared emission ($|b| > 20\deg$).
No error bars are available for the OVRO point.
To factor out variations in dust intensity $\Delta I_{100}$
over different patches of sky,
the figure plots the frequency dependence
of the correlation coefficient $\alpha$ instead.}
\label{alpha_vs_freq}
\end{figure}
% --------------------------------------------------------------------------

The problem is particularly acute for free-free emission:
lacking an accurate template for the spatial distribution,
estimation of free-free emission in microwave data
requires a pixel-by-pixel frequency decomposition
which significantly reduces the signal to noise ratio
of the desired cosmological signal.
Recently,
Kogut et al.\ (1996a,b)
proposed using infrared emission from diffuse cirrus
as a tracer of the ionized gas responsible for free-free emission,
\begin{equation}
\Delta T_{\rm ff} ~= ~\alpha ~\Delta I_{100}
\label{scaling_eq}
\end{equation}
where
$\Delta T_{\rm ff}$ is the fluctuation in microwave antenna temperature
from free-free emission, 
$\Delta I_{100}$ is the fluctuation in dust continuum intensity
at wavelength 100 $\mu$m,
and the coefficient $\alpha$ converts from units MJy/sr 
to $\mu$K antenna temperature.
Cross-correlation of the COBE
Differential Microwave Radiometer (DMR)
maps at 31.5, 53, and 90 GHz
with the Diffuse Infrared Background Experiment (DIRBE)
far-infrared maps at 100, 140, and 240 $\mu$m
shows statistically significant emission in each microwave map
whose spatial distribution
on angular scales above 7\deg
~is traced by the far-infrared dust emission.
The frequency dependence of this emission,
rising sharply at long wavelengths,
is inconsistent with the expected microwave dust emission
(Fig. \ref{alpha_vs_freq}).
A 2-component fit of the correlated COBE data 
to a model with dust plus radio emission
\begin{equation}
\Delta T_A ~= 
~\Delta T_{\rm dust} \left( \frac{\nu}{\nu_0} \right)^{\beta_{\rm dust}}
 ~+ ~\Delta T_{\rm radio} \left( \frac{\nu}{\nu_0} \right)^{\beta_{\rm radio}}
\label{beta_eq}
\end{equation}
with dust emissivity $1.5 < \beta_{\rm dust} < 2$
yields spectral index
$\beta_{\rm radio} = -2.1^{+0.6}_{-0.8}$
for the unknown component,
strongly suggestive of free-free emission ($\beta_{\rm ff} = -2.15$).
At high latitudes, 
spatial fluctuations in the diffuse synchrotron emission
are uncorrelated with dust emission,
leaving free-free emission (thermal bremsstrahlung)
from the WIM as seemingly the only plausible alternative.

The amplitude of the correlated component
agrees well with independent estimates of free-free emission
derived solely from its frequency dependence.
We estimate the amplitude of the total free-free emission
(including any uncorrelated component)
by analyzing a linear combination of the DMR maps
designed to be sensitive to free-free emission,
cancel emission with a CMB spectrum, and minimize instrument noise:
\begin{equation}
T_{\rm ff} =
  0.37 \times \frac{1}{2}(T^\prime_{\rm 31A} \pm T^\prime_{\rm 31B})
+ 0.02 \times \frac{1}{2}(T^\prime_{\rm 53A} \pm T^\prime_{\rm 53B})
- 0.47 \times \frac{1}{2}(T^\prime_{\rm 90A} \pm T^\prime_{\rm 90B}),
\label{ff_eq}
\end{equation}
where $T^\prime$ is the antenna temperature in each DMR channel
after subtracting synchrotron and dust emission
(Kogut et al.\ 1996b).
We smooth the maps with a 7\deg ~FWHM Gaussian
to further reduce the effects of noise,
remove a fitted monopole and dipole, and
compare the variance of the (A+B)/2 sum map to the (A-B)/2 difference map.
We obtain an estimate for the fluctuations
in free-free antenna temperature at 53 GHz from all sources,
$\Delta T_{\rm ff} = 5.2 \pm 4.2 ~\mu$K.
This value compares well with the correlated component
at the same 10\deg ~effective smoothing,
$\Delta T_{\rm radio} = 6.8 \pm 1.6 ~\mu$K
from Eq. \ref{beta_eq}.

% --------------- Figure 2: Spectral index chi-squared --------------- 
\begin{figure}
\plotone{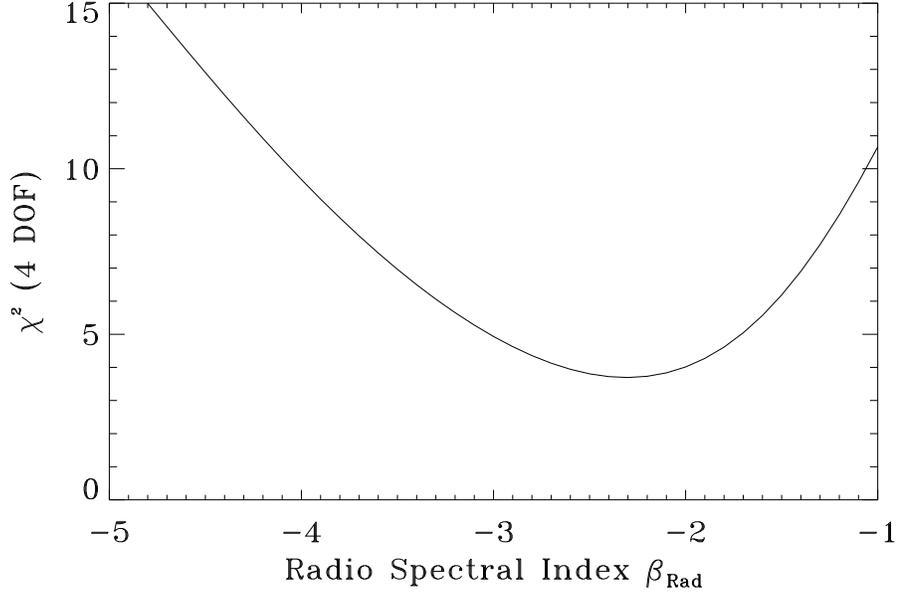}
\caption{Results of fitting all microwave data
from Figure \ref{alpha_vs_freq} to dust emission
plus emission with unknown spectral index $\beta_{{\rm radio}}$
(Eq. \ref{beta_eq}). }
\label{chi_sq_fig}
\end{figure}
% --------------------------------------------------------------------------

\section{Evidence for Anomalous Emission}

Two questions of interest for cosmological observations are
the extent to which infrared dust
reliably traces high-latitude radio emission,
and whether the correlation depends significantly on angular scale.
The detection of correlation 
between dust and ionized gas at high latitudes 
is consistent with the correlation observed 
between dust and free-free emission
along the Galactic plane
(Broadbent, Haslam, \& Osborne 1989),
and supports a picture
in which free-free emission from the WIM
is traced by continuum emission from warm (19 K) dust,
with scaling
$\alpha = 4.4 \pm 0.7 \left( \frac{\nu}{\rm 53 ~GHz} \right)^{-2.15} 
~\mu{\rm K} ~{\rm MJy}^{-1} ~{\rm sr}$
between microwave emission $\Delta T_{\rm ff}$
and dust emission $\Delta I_{100}$ at 100 $\mu$m wavelength.

A number of recent microwave observations
show that the observed correlation
between radio and far-IR emission
is ubiquitous
and not strongly dependent on angular scale.
Figure \ref{chi_sq_fig}
shows the spectral index
of the ``radio'' component
of the correlated emission
(Eq. \ref{beta_eq}),
derived from all current microwave detections
(DMR: Kogut et al.\ 1996b;
19 GHz: de Oliveira-Costa et al.\ 1998;
Saskatoon: de Oliveira-Costa et al.\ 1997; 
MAX5: Lim et al.\ 1996;
OVRO: Leitch et al.\ 1997).
The 95\% confidence interval
$ -3.6 < \beta_{\rm radio} < -1.3 $
is consistent with free-free emission
over a frequency range 14 -- 100 GHz.

When other tracers of the diffuse ionized gas are considered,
the picture becomes significantly more complicated.
H$\alpha$ emission is a widely used tracer of the 
diffuse gas in the WIM.
The intensity of optically thin H$\alpha$ emission 
is given by
\begin{equation}
I(R) =  0.44 {\rm EM} \left( \frac{T_e}{8000 ~{\rm K}} \right)^{-0.5}
        \left[ 1 - 0.34 \ln( \frac{T_e}{8000 ~{\rm K}} ) \right]
\label{rayleigh_eq}
\end{equation}
where ${\rm EM} = \int n_e^2 dl$
is the emission measure in cm$^{-6}$ pc
and the intensity units are Rayleighs
(1 R = $10^6$ photons / $4 \pi$ sr =
$2.42 \times 10^{-7}
~\rm{ergs~cm}^{-2} ~{\rm s}^{-1} ~{\rm sr}^{-1}$ at H$\alpha$).
Electron-ion collisions in the same gas
give rise to microwave free-free emission,
with antenna temperature
\begin{equation}
T_A = 0.83 Z^2 {\rm EM} ~
     \frac{ [1 + 0.23 \ln(T_e/8000~{\rm K})
        - 0.15 \ln(Z) - 0.15 \ln(\nu/53~{\rm GHz}) ] }
          { (\nu / 53~{\rm GHz})^2 ~(T_e/8000~{\rm K})^{1/2} }
\label{ff_eqn}
\end{equation}
in $\mu$K units, where $Z$ is the net charge of the ion
(Oster 1961; Bennett et al.\ 1992).
The WIM is characterized by electron density
$n_e \sim 0.1 ~{\rm cm}^{-3}$ 
and temperature $T_e \sim 8000$ K
(Reynolds 1990).  
H$\alpha$ emission should thus trace free-free emission as
\begin{equation}
\Delta T_{\rm ff} ~= ~a ~\Delta I_{{\rm H}\alpha},
\label{halpha_eq}
\end{equation}
with coefficient
$ a \approx 1.9 ~\left( \frac{\nu}{{\rm 53 ~GHz}} \right)^{-2.15}
\mu$K/R
(Bennett et al.\ 1992).

\begin{figure}
\plotone{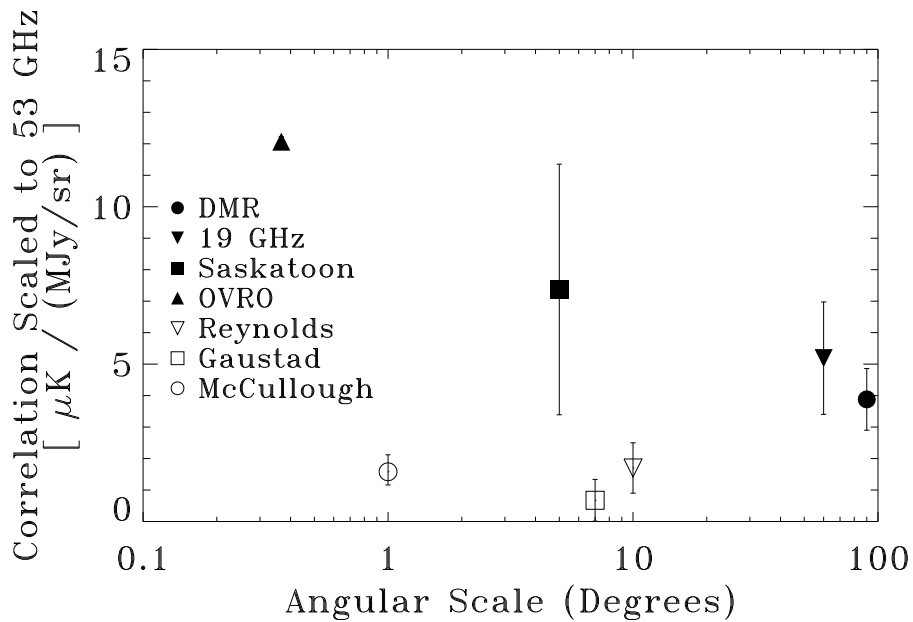}
\caption{Comparison of the correlation between
100 $\mu$m dust emission and
either H$\alpha$ or microwave data.
Microwave data use solid symbols
while H$\alpha$ results use open symbols.
Microwave values lie systematically higher than the H$\alpha$ results.
There is no apparent dependence on angular scale.}
\label{halpha_fig}
\end{figure}

If we assume that the observed correlation 
between microwave and far-IR maps
results from free-free emission,
then a similar correlation should exist 
between H$\alpha$ and dust emission.
Several authors have found marginal correlations
between H$\alpha$ maps and the IRAS or DIRBE 100 $\mu$m maps
(McCullough 1997,
Kogut 1997).
Figure \ref{halpha_fig} shows the correlation coefficient $\alpha$
between the far-IR dust continuum and
H$\alpha$ or microwave data,
where we have converted H$\alpha$ values from Rayleighs to $\mu$K
using Eq. \ref{halpha_eq}.
Both sets of data are positively correlated
with the far-IR dust emission,
but the microwave results appear systematically higher
than the H$\alpha$ values.

% --------------- Figure 3: H-alpha and microwave comparison --------------- 
% --------------------------------------------------------------------------

There are no full-sky maps of H$\alpha$ emission
to compare to the full-sky COBE data.
Several small H$\alpha$ maps exist
(Reynolds 1980,
Gaustad et al.\ 1996,
Simonetti et al.\ 1996,
McCullough 1997),
but only one field (the north celestial pole)
has been mapped both in H$\alpha$ and microwaves.
There, the small variation in H$\alpha$ intensity
appears inadequate to explain the
presumed free-free emission
by a factor 3--10,
albeit with substantial uncertainties
(Simonetti et al.\ 1996,
Leitch et al. 1997,
de Oliveira et al.\ 1997).
Leitch et al.\ (1997) argue 
that energetics preclude
solutions that reduce H$\alpha$ emission
by raising the gas temperature
and suggest that alternate explanations
are required for the ``anomalous'' microwave emission.

We can quantify this discrepancy using simple statistics.
If the microwave and H$\alpha$ results trace the same diffuse gas,
the values in Figure \ref{halpha_fig} should be drawn from a single
parent population.
Table \ref{chi_sq_table}
shows the $\chi^2$ derived from fitting a weighted mean
to the full data set
and two independent sub-sets
composed of the H$\alpha$ and microwave data, respectively.
The full data set has $\chi^2$ of 13.4 for 7 degrees of freedom,
improbable for normally-distributed data.
If we separate the data into microwave or H$\alpha$ results,
each set is internally consistent:
by adding a single degree of freedom 
(breaking the data into 2 subsets)
we reduce the total $\chi^2$ from 13.4 to 4.8,
suggesting that the data do, in fact,
belong to two separate populations.

% --------------- Table 1: Mean Correlation and Chi-Squared --------------- 
\begin{table}[t]
\caption{\label{chi_sq_table}Mean Correlation Coefficients at 53 GHz}
\begin{tabular}{l c c}
\tableline
Model & Fitted Correlation & $\chi^2 / {\rm DOF}$ \\
      & ($\mu$K / MJy sr)  &  \\
\tableline
All Data	&	2.6 $\pm$ 0.4	&	13.4 / 7 \\
Microwave Only	&	4.4 $\pm$ 0.7	&	 3.4 / 4 \\
H$\alpha$ Only	&	1.3 $\pm$ 0.4	&	 1.4 / 2 \\
\tableline
\tableline
\end{tabular}
\end{table}
% --------------------------------------------------------------------------

\section{Spinning Dust}

\begin{figure}
\plotfiddle{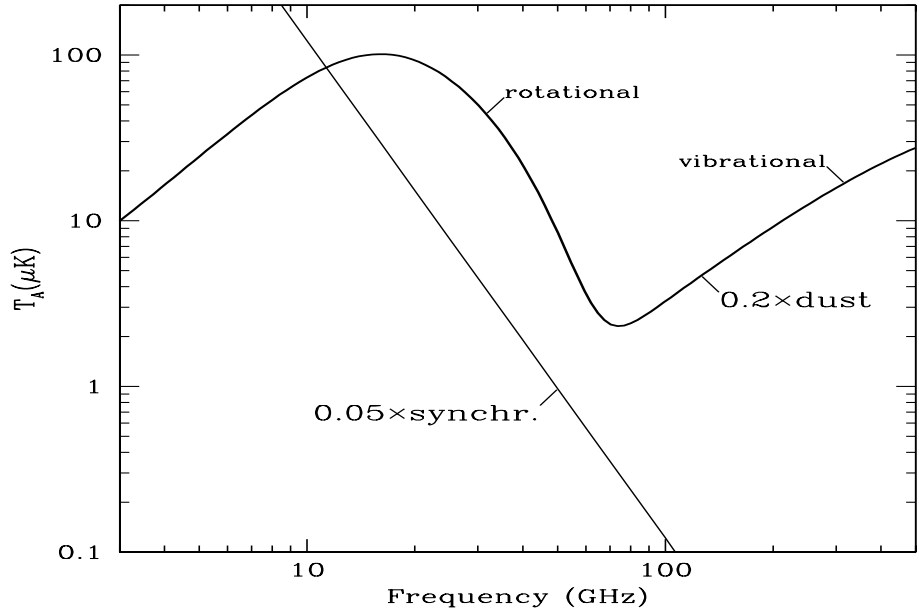}{3.0in}{0}{60}{40}{-200}{-56}
\caption{Predicted spectrum of electric dipole emission
from a population of small spinning dust grains
(from Draine \& Lazarian 1998a).}
\label{dust_spectra}
\end{figure}

Many experiments show microwave emission
correlated with far-IR dust emission,
with spectral index 
suggestive of free-free emission
but lacking the strong H$\alpha$ signal
expected on physical grounds.
If this is not free-free emission, what could it be?
Recent work suggests that the ``anomalous'' microwave emission
could originate from the electric dipole radiation
of very small ($N < 10^3$ atoms) spinning dust grains
(Ferrara \& Dettmar 1994;
Draine \& Lazarian 1998a,b).
A key feature of the spinning dust model 
is the microwave emission spectrum,
which peaks between 10 and 50 GHz
depending on the size distribution of the dust grains.
Figure \ref{dust_spectra} shows a typical spectrum,
with grain size ($N < 150$ atoms) chosen
to agree with the OVRO detection at 14.5 GHz.

Identification of ``anomalous'' microwave emission
as the signature of spinning dust
has several advantages.
It provides a natural explanation for the
observed correlation between emission components
at microwave and far-IR wavelengths:
both components results from the {\it same} 
physical dust distribution
and should be highly correlated at all angular scales.
It explains the lack of a strong H$\alpha$ signal
without invoking energetically questionable gas temperature.
Finally, since the emission spectrum falls sharply 
at wavelengths below 10 GHz, 
it allows detectable ``anomalous'' signals 
in the window 30--100 GHz
without violating limits from radio surveys
at 1420 and 408 MHZ
(Reich \& Reich 1988,
Haslam et al.\ 1981).

The identification of ``anomalous'' microwave emission
with spinning dust,
while suggestive,
is not without problems.
The spectral index $\beta_{\rm spin}$ (in antenna temperature) 
of the calculated dust emission
over the frequency range
19--53 GHz
where the most sensitive detections occur
is between -3.3 and -4,
depending on the assumed size distribution
of the dust grains
(Draine \& Lazarian 1998b).
Such a steep spectral index
lies 1.6 to 2.6 standard deviations
from the best fitted index $\beta_{\rm radio} = -2.2^{+0.5}_{-0.7}$
(Eq. \ref{beta_eq}).
Furthermore,
while long-wavelength limits are weak,
there is as yet no confirmation
of the predicted downturn
at frequencies below 20 GHz.

\section{Discussion}

A large body of observational evidence
demonstrates that at least one component of microwave emission
is traced by far-IR dust emission.
This correlation is ubiquitous on the sky
and independent of angular scale.
As such, the corresponding signal in microwave anisotropy maps
can be removed by correlating independently
each microwave frequency channel 
with a template dust map.
The amplitude at microwave frequencies
is small enough
that residual uncertainties,
after subtraction,
should not be a major source of error
for cosmological applications.\footnote{
If the correlated emission is in fact dominated by spinning dust,
the partial alignment of the grain spin axis
with Galactic magnetic fields
may be a source of confusion for measurements of the CMB polarization.}
%
%

% --------------- Figure 4: Spinning dust emission spectra --------------- 
% --------------------------------------------------------------------------

The source of the correlated emission
is still an open question
with interesting astrophysical implications.
The spectral index, within broad limits,
is consistent at the 95\% confidence level
with either free-free emission
or emission from spinning dust grains.
From physical arguments,
it seems likely that {\it both} emission mechanisms
may contribute a detectable signal
at microwave frequencies.
Both microwave and H$\alpha$ emission
are positively correlated with far-IR dust emission.
Assuming two separate emission mechanisms,
microwave data contain contributions from both free-free 
and spinning dust,
while H$\alpha$ results trace only the free-free component.
We may thus use the data in Table \ref{chi_sq_table}
to obtain separate free-free and dust normalizations at 53 GHz.
Averaged over the full sky,
emission from spinning dust is 
$2.5 \pm 1.1$ times as bright as free-free emission.

Substantial progress remains to be made in the identification
of microwave foregrounds.
A key discriminant between free-free and spinning dust models
is the spectrum below 20 GHz
where emission from spinning dust falls rapidly.
A measurement of the correlated emission
at a frequency below 10 GHz
would provide a definitive test of the two models.

Astrophysical modelling of microwave emission
as a probe of physical conditions in the interstellar medium
requires a clean separation of individual emission components.
Current efforts along these lines are hampered by the
lack of high signal-to-noise ratio maps
at microwave, H$\alpha$, and far-IR frequencies
in common regions of the sky.
H$\alpha$ maps from the Wisconsin H-Alpha Mapper
(Tufte, Reynolds, \& Haffner 1998),
in combination with microwave data
from the next generation balloon and satellite experiments,
should allow a full set of microwave--H$\alpha$--dust correlations
to separate emission from different phases of the ISM.

Finally, sensitive microwave maps with substantial sky coverage
will test for significant variation in the correlated emission
from point to point on the sky.
While there is no evidence for spatial variation in
the correlation,
existing full-sky data sets (the COBE and 19 GHz maps)
are too noisy to test for variation in the correlation coefficient
on angular scales below 60\deg.
Two of the three detections from small patches of sky
(OVRO and MAX5) show appreciably larger correlations 
than the full-sky COBE results,
but with substantial uncertainties.
How anomalous are these anomalous results?
The MAP and Planck satellite data should demonstrate whether
full-sky average correlation is physically meaningful,
or whether a more complicated model is required.

% Th-th-th-that's all, folks!
\end{document}